\documentclass[english,letterpaper,twocolumn,showpacs,prl,aps]{revtex4}
\usepackage{graphicx}
\usepackage{amssymb}

\makeatletter

\large  

\input epsf

\makeatother

\usepackage{babel}
\makeatother
\begin{document}

\title{Fermion condensation around a Coulomb impurity in a Weyl semimetal and  in a narrow band gap semiconductor as manifestations of the Landau zero-charge problem}

\author{Eugene B. Kolomeisky$^{1}$ and Joseph P. Straley$^{2}$}

\affiliation
{$^{1}$Department of Physics, University of Virginia, P. O. Box 400714,
Charlottesville, Virginia 22904-4714, USA\\
$^{2}$Department of Physics and Astronomy, University of Kentucky,
Lexington, Kentucky 40506-0055, USA}

\begin{abstract}
A Coulomb impurity placed in an undoped Weyl semimetal spontaneously surrounds itself with a cloud of condensed Weyl fermions.  We find that the ground-state of this system exhibits an experimentally accessible Landau zero-charge effect:  the fermion condensate completely screens out the impurity charge.  In a narrow band gap semiconducor this effect manifests itself in the near universality of observable charge of a highly-charged recombination center.    
\end{abstract}


\maketitle

The Dirac equation for an electron in the field of a point charge $Ze$ in vacuum becomes meaningless for $Z>1/\alpha\approx137$ ($\alpha=e^{2}/\hbar c$ is the fine structure constant) because it predicts an imaginary ground-state energy \cite{LL4}.   For a nucleus with $Z_{c}\approx170$ (upshifted from $Z = 137$ as it is not a point particle) the total energy of the production of an electron-positron pair vanishes and the vacuum becomes unstable with respect to pair creation;  the positron repelled by the nucleus escapes to infinity while the electron remains near the nucleus \cite{ZP}.  For $Z>Z_{c}$ vacuum electrons condense around the nucleus within a shell partially screening the nuclear charge.  As a result, the observable dimensionless nuclear charge as seen at large distances is less than its bare value of $Z$ \cite{numerical, MVP}.   The peculiarity of this system is that its ground state, the vacuum, is charged.

Here we observe that the prediction of vacuum electron condensation in quantum electrodynamics (QED) and of "charged vacuum" can be tested in performable experiments involving condensed matter systems, both presently available and those that will become available in the near future.  Our primary example is that of semiconductors whose physics is known to mimic, to some extent, QED \cite{ZP}.  The excitation of an electron-hole pair parallels the creation of an electron-positron pair in QED, with the band gap imitating the combined rest energy of the particles.  There is also a counterpart to  a $Z\gtrsim 137$ nucleus in condensed matter physics.  In his study of the impurity states in semiconductors Keldysh \cite{Keldysh} noted that the effective mass approximation \cite{LL9}, while successful in describing shallow impurity states,  fails to explain deep states whose binding energy is comparable with the band gap.  Such states are formed near highly-charged impurities and (in contrast to their shallow counterparts) they cannot be associated with either conduction or valence bands.  The experiment presented another puzzle:   some highly charged impurities acted as very efficient recombination centers but an explanation why that was the case was lacking.  Keldysh showed that experimental findings can be understood in a two-band approximation (well obeyed in narrow band gap semiconductors (NBGS) of the $InSb$ type) where the low-energy electron-hole dispersion law is relativistic \cite{Keldysh,Kane}
\begin{equation}
\label{dispersion}
\varepsilon(\textbf{p})=\pm \sqrt{(\Delta/2)^{2}+v^2 p^{2}}
\end{equation}            
where $\Delta$ is the energy band gap and $v$ is the velocity of a high-momentum particle analogous to the speed of light in vacuum $c$;  in NBGS we have $v\approx 4.3 \times 10^{-3}c$ \cite{Zawadzki}.  Then determination of the impurity states reduces to solving the Dirac equation in the field of a charge $Ze$ screened by the dielectric constant $\epsilon$ of the semiconductor.  In view of the $Z\alpha>1$ peculiarity of the Dirac equation (now $\alpha=e^{2}/\hbar v\epsilon$), Keldysh concluded that for $Z\alpha < 1$ the impurity states are given by the known solution to the Dirac equation \cite{LL4} while the anomalous $Z\alpha>1$ case describes a recombination center.  

The recombination center is clearly a semiconductor counterpart of the $Z\gtrsim137$ nucleus.  However,  the semiconductor equivalent of the fine structure constant is $\alpha=e^{2}/\hbar v \epsilon \approx 1.7/\epsilon$.  Since we now have $\epsilon \simeq 10$ \cite{Kittel}, the corresponding $\alpha$ is an order of magnitude larger than its QED counterpart.  Thus the characteristic $Z=1/\alpha$ of the onset of the electron condensation is about $10$.  Surely a $Z\gtrsim10$ recombination center is a more common object than a $Z\gtrsim137$ nucleus.  

In addition to making it possible to study the regime of large effective fine structure constant, condensed matter systems also offer possibilities that cannot be realized in QED.  Indeed, over forty years ago Abrikosov and Beneslavski\u{i} \cite{AB} predicted the existence of semimetals having points in the Brillouin zone where the valence and conduction bands meet with a dispersion law that is linear in the momentum.  This is the $\Delta =0$ case of Eq.(\ref{dispersion}). Such systems, nowadays called  Weyl semimetals (WS), are likely to be realized in doped silver chalcogenides $Ag_{2+\delta}Se$ and $Ag_{2+\delta}Te$ \cite{silver}, pyrochlore iridates $\mathcal{A}_{2}Ir_{2}O_{7}$ (where $\mathcal{A}$ is Yttrium or a lanthanide) \cite{pyro}, and in  topological insulator multilayer structures \cite{topo}.  The zero energy gap of a WS implies that there is no threshold to creation of the electron-hole pairs.  Thus condensation of Weyl fermions around a Coulomb impurity is spontaneous.  The physically related problem of impurity screening in graphene has been considered elsewhere \cite{graphene} (that problem is mathematically different from what we discuss, because it is a two-dimensional semimetal embedded in a three-dimensional space).    

Below we will determine the ground-state properties of NBGS (including its WS limit) in the presence of a Coulomb impurity, as a function of $Z$ and $\alpha$.  At modest $Z$ the electrons of the "vacuum" (valence band) condense around the impurity while the holes leave the physical picture; the properties of the electron cloud vary with $Z$ and $\alpha$ and are determined by the interplay of attraction to the impurity (promoting electron condensation), and electron-electron repulsion combined with the Pauli principle (limiting the condensation).  The QED analysis of the physical properties of the vacuum electron cloud was carried out in two limits:

(i) $Z$ close to $Z_{c}$, where there are very few condensed electrons and the single-particle picture is a good starting point \cite{ZP}, and

(ii) $Z\gg Z_{c}$, where the number of condensed electrons is large and the electron-electron interactions cannot be ignored \cite{numerical,MVP}.

The goal of this paper is to demonstrate that the physics of the charged vacuum in the $Z\gg Z_{c}$ limit exhibits large degree of universality manifesting itself in a nearly-universal observable impurity charge;  its detection is within experimental capabilities.  Although below we adopt the semiconductor language, our findings are also applicable in the QED implementation of the problem.  Specifically, our central conclusion that in the gapless $\Delta=0$ case the large distance character of screening is formally identical to that occurring in the Landau "zero-charge" problem \cite{LL4} was missed in previous studies \cite{MVP}.   

Since for $Z\gg Z_{c}$ a large number of condensed electrons is present, the properties of the system consisting of the impurity and its interacting cloud of electrons can be understood semiclassically with the help of the Thomas-Fermi (TF) theory \cite{numerical,MVP} which becomes asymptotically exact in the $Z\rightarrow \infty$ limit \cite{Spruch}.  The main object of the TF theory is a physical electrostatic potential $\varphi(\textbf{r})$ felt by an electron that is due to both the electrostatic potential of the impurity $\varphi_{ext}(\textbf{r})$ and that of the condensed electrons characterized by the number density $n(\textbf{r})$:
\begin{equation}
\label{definition_potential}
\varphi(\textbf{r})=\varphi_{ext}(\textbf{r})-\frac{e}{\epsilon}\int \frac{n(\textbf{r}')dV'}{|\textbf{r}-\textbf{r}'|}
\end{equation}
The external potential $\varphi_{ext}(\textbf{r})$ is a pseudopotential that represents the perturbation of the system caused by the impurity; even though $\varphi_{ext}$ is not entirely of electrostatic origin, we will define $\triangle \varphi_{ext} = -4\pi e n_{ext}/\epsilon$.  We assume that the impurity charge density $en_{ext}(\textbf{r})$ is spherically-symmetric and localized within a region of size $a\simeq1nm$ so that for $r > a$ the potential $\varphi_{ext}(\textbf{r})$ reduces to a purely Coulomb form $\varphi_{ext}(r)= Ze/\epsilon r$ of a net charge $Ze$ within the impurity region.  Given $\varphi(\textbf{r})$, one can deduce that the electron number density $n(\textbf{r})$ is different from zero only in the region of space where the electron potential energy $-e\varphi(\textbf{r}) +\Delta/2$ drops below $-\Delta/2$, thus defining the "vacuum" shell where condensed electrons reside as
\begin{equation}
\label{shell_region}
e\varphi(\textbf{r})> \Delta,~~~n(\textbf{r})>0
\end{equation}    
The radius of the electron shell $R_{vac}$ is given by the equalities $e\varphi(R_{vac})=\Delta$, $n(R_{vac})=0$;  outside the shell we have $n=0$ and 
\begin{equation}
\label{observable_charge_definition}
\varphi=\frac{Q_{\infty}e}{\epsilon r},~r>R_{vac}=Q_{\infty}\frac{e^{2}}{\epsilon \Delta}\equiv \frac{aQ_{\infty}}{Z_{0}},~Z_{0}=\frac{\epsilon \Delta a}{e^{2}}
\end{equation}
where $Q_{\infty}<Z$ is the dimensionless observable impurity charge as seen at large distances from its center.  Continuity of the potential $\varphi$ across the shell boundary relates $R_{vac}$ and $Q_{\infty}$ as indicated in the last two steps in (\ref{observable_charge_definition}) meaning that we can speak of the shell size or the observable charge interchangeably.  The parameter $Z_{0}<Z_{c}$ gives the classical, $\alpha=\infty$, value of the critical charge of the onset of the electron condensation at the edge of the impurity region;  in NBGS with $\Delta\simeq0.1eV$ and $a\simeq 1nm$ we have $Z_{0}\simeq1$.  In WS ($Z_{0}=0$) we find $R_{vac}=\infty$, i.e. the screening cloud has infinite extent.

From the thermodynamical standpoint, electron ($e$) condensation in the field of a Coulomb impurity accompanied by escape of a hole ($h$) to infinity may be viewed as a "chemical reaction" $e+h\leftrightarrows 0$ (the ground state of the semiconductor is the "vacuum") \cite{LL9};  the condition of equilibrium for this reaction has the form
\begin{equation}
\label{chem_potentials}
\mu_{e}+\mu_{h}=0,\mu_{e}=\sqrt{(\Delta/2)^{2}+v^2 p_{F}^{2}}-e\varphi,\mu_{h}= \Delta/2
\end{equation}
where $\mu_{e}$ and $\mu_{h}$ are the chemical potentials of the electrons and holes, respectively, and $p_{F}(\textbf{r})=\hbar [6\pi^{2}n(\textbf{r})/g]^{1/3}$ is the Fermi momentum which we assume is a slowly varying function of position.  The parameter $g$ is the fermion degeneracy factor: $g=2$ in QED while in NBGS it is twice the number of Dirac valleys (\ref{dispersion}) within the first Brillouin zone; an isotropic valley-independent limiting velocity $v$ is assumed for simplicity.   In the WS case $g$ counts the number of Weyl points within the first Brillouin zone:  $g=24$ in pyrochlore iridates \cite{pyro} and  $g=2$  in a topological insulator multilayer \cite{topo}.  Then the condition (\ref{chem_potentials}) implies a relationship between the physical potential and the electron number density \cite{numerical,MVP}:
\begin{equation}
\label{n_of_phi}
n(\textbf{r})= \frac{\lambda}{4\pi} \bigg\{\frac{\epsilon \varphi(\textbf{r})}{e}\frac{\epsilon}{e^{2}}[e\varphi(\textbf{r})-\Delta]\bigg\}^{3/2}, ~~\lambda = \frac{2g\alpha^{3}}{3\pi}
\end{equation} 
where $\lambda$ is a dimensionless parameter that characterizes the relative strength of electron-electron interactions and zero-point motion.  It can be as small as $1.65 \times 10^{-7}$ (QED) or as "large" as $10^{-3}$ (NBGS).  

Applying the Laplacian operator to both sides of Eq.(\ref{definition_potential}) and using (\ref{n_of_phi})  we find the relativistic TF equation 
\begin{equation}
\label{diff_eq_phi}
\nabla^{2} \left (\frac{\epsilon \varphi}{e}\right ) =-4\pi n_{ext} +\lambda \bigg\{\frac{\epsilon \varphi}{e} \frac{\epsilon}{e^{2}}(e\varphi-\Delta)\bigg\}^{3/2}
\end{equation}
that was investigated in QED \cite{numerical,MVP}.  The source term is localized and its specific form is not very important; we can take $n_{ext}$ to have the constant value $3Z/4\pi a^3$.  A numerical solution to (\ref{diff_eq_phi}) is shown in the Figure, where we additionally displayed the charge $Q(r)$ within a sphere of radius $r$ as an indicator of the strength of screening.  In agreement with previous analysis \cite{numerical,MVP} the screening effect of condensed electrons becomes noticeable for $\lambda Z^{2}\gtrsim1$.  
\begin{figure}
\includegraphics[width=1.0\columnwidth, keepaspectratio]{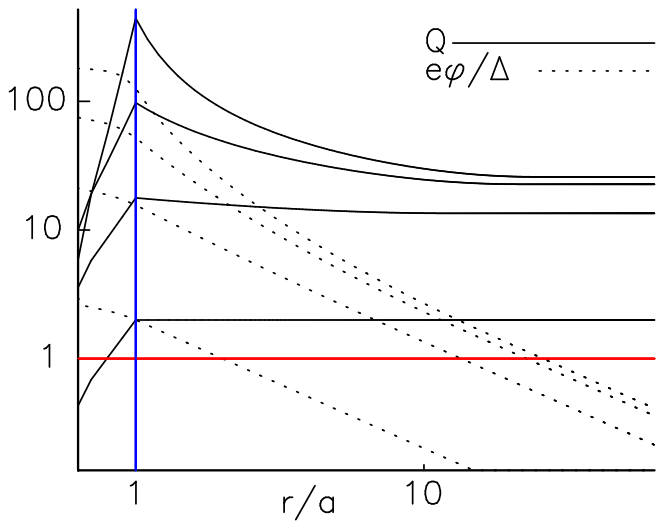} 
\caption{(Color online) Potential $\varphi$ (\ref{definition_potential}) and charge $Q(r)$ within a sphere of radius $r$ (\ref{charge_connection}) as functions of distance (double logarithmic representation).  The source region is $r/a < 1$.  The gap value $\Delta$ is indicated by the horizontal line; the electron cloud is limited to the region where $e\varphi >  \Delta$, which defines $R_{vac}$.  The curves are drawn for $Z = 2, 20, 200, 2000$, $\lambda = 0.001$, and $Z_{0}=1$ (other values of $Z_{0}$ are equivalent to a rescaling $Z\rightarrow Z/Z_{0}$ and $\lambda\rightarrow\lambda Z_{0}^{2}$).  For large $Z$, $R_{vac}$ approaches a $Z$-independent limit, indicating that $Q(r)$ tends to an upper bound $Q_{\infty}$.}
\end{figure} 
Remarkably, for $Z\gg1$ there exists a $Z$-independent limit on $R_{vac}$ and $Q_{\infty}$.     To understand the form of numerical solution in the strong screening regime $\lambda Z^{2}\gg1$, it is useful to start with the WS case, $\Delta=0$, when Eq.(\ref{diff_eq_phi}) simplifies to
\begin{equation}
\label{diff_eq_phi_m=0}
\nabla^{2} \left (\frac{\epsilon \varphi}{e}\right ) =-4\pi n_{ext} +\lambda \left ( \frac{\epsilon \varphi}{e}\right )^{3}
\end{equation}  
In the strong screening $\lambda Z^{2}\rightarrow\infty$ case the solution to Eq. (\ref{diff_eq_phi_m=0}) is given by the zero of its right-hand side: 
\begin{equation}
\label{leading_strong_screening_n_phi}
n=n_{0}= n_{ext}, ~~~~~~\varphi= \varphi_{0}=\frac{e}{\epsilon}\left (\frac{4\pi n_{ext}}{\lambda}\right )^{1/3}
\end{equation}
We see that to leading $\lambda Z^{2}=\infty$ order the screening is complete with zero electric field everywhere and constant potential $\varphi_{0}$ inside the impurity region \cite{numerical,MVP}.  In order to further improve on (\ref{leading_strong_screening_n_phi}), within the impurity region we substitute $\varphi=\varphi_{0}(1-f)$, $0\leqslant f\ll1$, into Eq.(\ref{leading_strong_screening_n_phi}) and linearize about $\varphi=\varphi_{0}$:  
\begin{equation}
\label{linearized_TF_inside}
\nabla^{2} f-\kappa^{2}f=0, ~~~~~\kappa^{2} a^{2}=3^{5/3}(\lambda Z^{2})^{1/3}\gg1
\end{equation}
This approximation parallels the TF theory of screening in a Fermi gas \cite{Kittel}.  We observe that inside the impurity region the screening response is characterized by the TF screening length $\kappa^{-1}\simeq a (\lambda Z^{2})^{-1/6}\ll a$:  it is a length scale over which $f$ drops to practically zero from a value it assumes on the impurity boundary.  This is also the width of the region adjacent to the impurity boundary where uncompensated charge is localized \cite{MVP}.  From here the net charge inside the impurity region can be estimated as $(Z/a^{3})a(\lambda Z^{2})^{-1/6}a^{2}=Z(\lambda Z^{2})^{-1/6}\ll Z$ \cite{MVP}.  The crossing of the charge curves for $Z=200$ and $Z=2000$ at small $r$ shown in Figure is a direct consequence of screening: the TF screening length $\kappa^{-1}$ is smaller for $Z=2000$ than $Z=200$, so that the screening at the central region is more complete in the former case.  

The solution to Eq.(\ref{linearized_TF_inside}) that is finite at the origin has the form $f\propto \sinh(\kappa r)/r$, so that for $r\leqslant a$, 
\begin{equation}
\label{potential_inside}
\frac{\epsilon \varphi}{e}=\frac{\epsilon \varphi_{0}}{e}(1-f)=\frac{9Z}{\kappa^{2}a^{3}}\left (1- A\frac{a\sinh(\kappa r)}{r\sinh(\kappa a)}\right )
\end{equation}          
where $A$ is a constant assumed to be much smaller than unity to justify linearization approximation (\ref{linearized_TF_inside}).

Outside of the source one has to look at the full non-linear equation (\ref{diff_eq_phi_m=0}) whose solution is sought in the form 
\begin{equation}
\label{def_zeta}
\frac{\epsilon \varphi(r)}{e}= \frac{1}{r}\zeta \left (\frac{r}{a}\right )  
\end{equation}
where, via Gauss's theorem, the function $\zeta$ is related to the charge $Q(r)$ within a sphere of radius $r$ as:
\begin{equation}
\label{charge_connection}
Q(r)=-r^{2}\frac{\partial(\epsilon\varphi/e)}{\partial r}=\zeta(l)-\zeta'(l),~~~l=\ln\frac{r}{a}
\end{equation} 
Substituting (\ref{def_zeta}) into (\ref{diff_eq_phi_m=0}) for $r>a$ we obtain the equation
\begin{equation}
\label{diff_eq_zeta_of_x}
\zeta''(l)-\zeta'(l)=\lambda\zeta^{3} .
\end{equation}
For $l=\ln(r/a)\ll1$ we can neglect in Eq.(\ref{diff_eq_zeta_of_x}) the first-order derivative term $\zeta'(l)$ compared to $\zeta''(l)$;  then $Q(r)\approx -\zeta'(l)$.  The solution to (\ref{diff_eq_zeta_of_x}) in this limit is
\begin{equation}
\label{zeta_near_boundary}
\zeta_{1}(l)=\sqrt{\frac{2}{\lambda}}\frac{1}{l+B}, ~~~0\leqslant l\ll 1
\end{equation}
Continuity of the potential and of the electric field at the impurity boundary $r=a$ determines the integration constants $A$ and $B$ in Eqs.(\ref{potential_inside}) and (\ref{zeta_near_boundary}) to be
\begin{equation}
\label{int_constants}
A\approx 0.2374, ~~~B\approx \frac{1}{0.3113 \kappa a}\simeq (\lambda Z^{2})^{-1/6}\ll 1
\end{equation}
The solution (\ref{potential_inside}), (\ref{def_zeta}) and (\ref{zeta_near_boundary}) also describes the NBGS case since the condition $e\varphi \gg \Delta$ necessary for transition from (\ref{diff_eq_phi}) to (\ref{diff_eq_phi_m=0}) holds.  Specifically, in the $Z\rightarrow \infty$ limit the parameter $B$ vanishes and the $Z$-dependence drops out of (\ref{zeta_near_boundary}).  The solution to the full TF equation (\ref{diff_eq_phi}) for $r>a$ then satisfies the singular boundary condition $\epsilon \varphi(r)/e\rightarrow \sqrt{2/\lambda}(r-a)^{-1}$ as $r\rightarrow a$ leading to the numerically observed $Z$-independent limit on $R_{vac}$ and $Q_{\infty}$.    

In the strong-screening limit the solution to the full Eq.(\ref{diff_eq_zeta_of_x}) has the form $\zeta(\lambda, l)=(2/\lambda)^{1/2}y(l)$ where $y(l)$ is a parameter free universal function such as $y(l\rightarrow 0)\rightarrow l^{-1}$.  The latter behavior is no longer an accurate representation of the true dependence $y(l)$ past $l\simeq1$.  Therefore the solution (\ref{zeta_near_boundary}) is only applicable up to a crossover scale $l=l^{*}\simeq1$, i.e. within several impurity radii.  Within this range the rescaled potential $\epsilon \varphi/e$ drops from a value of the order $\lambda^{-1/2}(\lambda Z^{2})^{1/6} a^{-1}$ at the impurity boundary to $\lambda^{-1/2}a^{-1}$ at the crossover scale $l^{*}$.  This explains the large slope of the potential and charge curves near the boundary that can be observed in the Figure.

For $l=\ln(r/a)\gg1$ we can neglect in Eq.(\ref{diff_eq_zeta_of_x}) the second-order derivative term $\zeta''(l)$ compared to $\zeta'(l)$;  then $Q(r)\approx \zeta(l)$ and  for \textit{arbitrary} screening strength Eq.(\ref{diff_eq_zeta_of_x}) acquires a form
\begin{equation}
\label{GL}
\frac{dQ}{dl}=-\lambda Q^{3}
\end{equation}
that is mathematically identical to the Gell-Mann-Low equation \cite{LL4} for the physical charge in QED reflecting the effects of vacuum polarization.   Eq.(\ref{GL}) exhibits the effect of "zero charge":  no matter what the "initial" value of $Q$ is, the system "flows" to the zero charge fixed point $Q=0$ as $l\rightarrow \infty$, i.e. the impurity charge has been completely screened.  In the strong-screening regime Eq.(\ref{GL}) is applicable at $l\gtrsim l^{*}\simeq 1$.  As a result the charge inside a sphere of radius $r>a^{*}=ae^{l^{*}}\gtrsim a$ will be given by 
\begin{equation}
\label{0_charge_solution}
Q^{2}(r)=\frac{Q^{*2}}{1+2\lambda Q^{*2}\ln(r/a^{*})}\rightarrow \frac{1}{2\lambda  \ln(r/a^{*})}
\end{equation}
where the integration constant $Q^{*}$ is the charge within a sphere of radius $a^{*}$.  Since $Q^{*}\simeq \zeta_{1}(l^{*})\simeq \lambda^{-1/2}\gg1$, the charge $Q(r)$ is given by the last representation in (\ref{0_charge_solution}) whose hallmark is \textit{near universality}: a weak logarithmic dependence on the source size $a\simeq a^{*}$ with universal amplitude.  We conclude that in the WS case, except for the immediate vicinity of the impurity boundary where Eq.(\ref{zeta_near_boundary}) applies, the solution to the screening problem is nearly-universal.  

In the NBGS case ($\Delta\neq 0$) the solution (\ref{0_charge_solution}) remains relevant at distances $r\ll R_{vac}$; however the true solution $Q(r)$ corresponding to (\ref{0_charge_solution}) merges smoothly with $Q_{\infty}$ at $R_{vac}$.  The latter can be estimated with logarithmic accuracy by equating $e \varphi$ to $\Delta$, with the results
\begin{equation}
\label{results}
Q_{\infty} \approx \frac {1}{\sqrt {2 \lambda \ln\Gamma}},~R_{vac} \approx a\frac{\Gamma}{\sqrt{\ln\Gamma}},~~~~~\Gamma= \frac{1}{Z_{0}\sqrt{2\lambda}}
\end{equation}
The condensed electron cloud is of finite spatial extent and the screening is incomplete;  the results (\ref{results}) are accurate provided $\ln \Gamma \gg1$.  For NBGS with $a\simeq 1nm$ impurity region ($Z_{0}\simeq1$) and $\lambda =10^{-3}$ we find $\ln\Gamma\approx3$.  This is not very large but sufficient to estimate the limiting charge as $Q_{\infty}\simeq(2\lambda)^{-1/2}\approx 22$, and the size of the vacuum shell $R_{vac}\simeq 22nm$, in agreement with numerical solution of the problem.  In the QED context we find $Q_{\infty}\simeq137^{3/2}\approx1600$.  

We note that making the substitution $Q^{*}\rightarrow Z$, $a^{*}\rightarrow a$ in Eq.(\ref{0_charge_solution}) gives an accurate interpolation formula that describes the regime of weak screening ($\lambda Z^{2}\ll1$) for all $r$.  The nearly universal limit of Eq.(\ref{0_charge_solution}) will be reached at distances $r\gtrsim a \exp(1/2\lambda Z^{2})$.  Applying to this solution the condition $e\varphi=\Delta$ defining the edge of the electron vacuum shell, we recover, with logarithmic accuracy, earlier results \cite{MVP} corresponding to the regime of weak screening in NBGS.   
 
This work was supported by US AFOSR Grant No. FA9550-11-1-0297.

\end{document}